\documentclass[aps,prb,twocolumn,longbibliography,superscriptaddress,10pt]{revtex4-2} %two column version
\usepackage[utf8]{inputenc}
\usepackage{amssymb,amsmath,amsfonts}
\usepackage{graphicx}
\usepackage{natbib}
\usepackage{multirow}
\usepackage{physics}
\usepackage{gensymb}
\usepackage{subfigure}
\usepackage{hyperref}
\usepackage[dvipsnames]{xcolor}
\usepackage{soul}
\usepackage{comment}
\usepackage[T1]{fontenc}
\renewcommand{\vec}[1]{\textbf{\textit{#1}}}

\newcommand{\orcidicon}[1]{\href{https://orcid.org/#1}{\includegraphics[height=\fontcharht\font`\B]{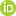}}}

\begin{document}

\title{Singularity theory of Weyl-point creation and annihilation}

\author{Gy\"orgy Frank\,\orcidicon{0000-0003-2129-2105}}

\affiliation{Department of Theoretical Physics, Institute of Physics, Budapest University of Technology and Economics, M\H{u}egyetem rkp. 3., H-1111 Budapest, Hungary}

\author{Gerg\H{o} Pint\'er}

\affiliation{Department of Theoretical Physics, Institute of Physics, Budapest University of Technology and Economics, M\H{u}egyetem rkp. 3., H-1111 Budapest, Hungary}

\author{Andr\'as P\'alyi}

\affiliation{Department of Theoretical Physics, Institute of Physics, Budapest University of Technology and Economics, M\H{u}egyetem rkp. 3., H-1111 Budapest, Hungary}

\affiliation{MTA-BME Quantum Dynamics and Correlations Research Group, M\"uegyetem rkp. 3., H-1111 Budapest, Hungary}

\begin{abstract}
Weyl points (WP) are robust spectral degeneracies, which can not be split by small perturbations, as they are protected by their non-zero topological charge. For larger perturbations, WPs can disappear via pairwise annihilation, where two oppositely charged WPs merge, and the resulting neutral degeneracy disappears. The neutral degeneracy is unstable, meaning that it requires the fine-tuning of the perturbation.  Fine-tuning of more than one parameter can lead to more exotic WP mergers. In this work, we reveal and analyze a fundamental connection of the WP mergers and singularity theory: phase boundary points of Weyl phase diagrams, i.e., control parameter values where Weyl point mergers happen, can be classified according to singularity classes of maps between manifolds of equal dimension. We demonstrate this connection on a Weyl--Josephson circuit where the merger of 4 WPs draw a swallowtail singularity, and in a random BdG Hamiltonian which reveal a rich pattern of fold lines and cusp points. Our results predict universal geometrical features of Weyl phase diagrams, and generalize naturally to creation and annihilation of Weyl points in electronic (phononic, magnonic, photonic, etc) band-structure models, where Weyl phase transitions can be triggered by control parameters such as mechanical strain.
\end{abstract}

\maketitle

\tableofcontents

\section{Introduction}

\begin{figure}
	\begin{center}
		\includegraphics[width=0.6\columnwidth]{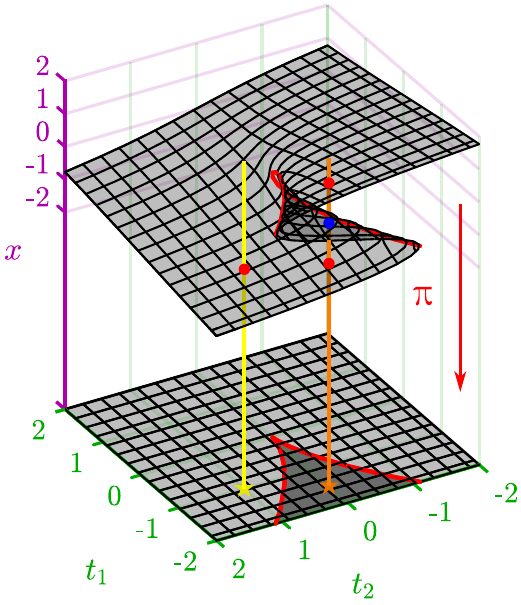}
  \end{center}
\caption{Fold and cusp singularities of the projection of a curved 2D manifold to a flat 2D manifold.
The Weyl points in the $n_\text p$-dimensional total parameter space of a physical system described by Hermitian matrices usually form an $m = n_\text p-3$-dimensional manifold. A minimal model of this Weyl-point manifold is illustrated here with a surface of dimension $m = 2$, parametrized by $(x,t_1,-x^3-t_1x)$ in the three-dimensional space of $(x,t_1, t_2)$. Separating the total parameter space into a 1D configurational ($x$) and 2D control ($t_1$,$t_2$) space correspond to a projection $\pi$. The number of Weyl points in the configurational space corresponding to a control parameter set $(t_1,t_2)$ is the number of pre-images $\#\pi ^{-1}(t_1,t_2)$ of the projection. The characteristic Weyl-point merger processes correspond to the singularities (see text) of the projection.}
\label{fig:projection}
\end{figure}

Singularity theory \cite{Arnold2012} provides a classification of singularities of mappings between manifolds. 
An instructive and easy-to-visualise example, where the dimension of both manifolds is $m=2$, is shown in Fig.~\ref{fig:projection}.
The source manifold is the curved surface embedded in the 3D space, the target manifold is a plane, and the mapping $\pi$ is the projection of the curved surface to the plane.
The singular points of this mapping are red points of the curved surface, i.e., those points that are mapped to the red points of the flat surface.

According to Whitney's theorem \cite{Whitney1955}, there are two classes of singular points in this setting: the \emph{fold} class, exemplified by the pre-images of the values forming the two red curves, and the \emph{cusp} (or \emph{pleat}) class, exemplified by the pre-image of the meeting point of the two red curves. 
In fact, Whitney's theorem asserts that for a \emph{generic} mapping between two two-dimensional (2D) manifolds, the singular points belong to one of these two classes. 
Further work from Mather \cite{Mather1969} has generalised this classification for mappings between higher-dimensional manifolds.
The classes of singular points (which, in technical terms, are left-right equivalence classes of map germs) are often referred to as \emph{singularities}.

Singularity theory (sometimes referred to as catastrophe theory \cite{Arnold1986}) is strongly interlinked with physics \cite{Stewart1981,Poston2014}, e.g., via applications in optics \cite{Berry1980,Zannotti2017}, seismology \cite{Brown1987}, molecular physics \cite{Kusmartsev1989,Krokidis1997}, band-structure theory \cite{Liu2014,Teramoto2017,Chandrasekaran2020,Yuan2020}, Hermitian and non-Hermitian quantum mechanics \cite{Kaufmann2012,Hu2023,Garmon2023}, and dynamical systems \cite{Zeeman1976}.

In particular, a recent work \cite{Hu2023,Garmon2023} discovered and analysed, both in theory and in experiment, a new link between singularity theory and physics. 
That work has revealed that the swallowtail singularity, characteristic of mappings between manifolds of dimension $m=3$, can appear as the phase diagram in the three-dimensional parameter space of the studied physical system, which is described by a parameter-dependent $3 \times 3$ non-Hermitian Hamiltonian matrix with a particular symmetry.

In this work, we show that the singularites classified by Whitney and Mather naturally appear in physical systems that are described by parameter-dependent Hermitian matrices -- a ubiquitous situation in quantum mechanics. 
We focus on the case when the number $n_\text p = 3 + m$ of parameters is greater than 3, and the parameters can be grouped into two groups: a group with 3 parameters, which we call the `configurational parameters', and another group with $m$ parameters, which we call the `control parameters'. 
This setting is relevant for many physical systems, e.g., (i) electronic (phononic, magnonic, photonic) band structure theory of 3D crystals, where the configurational space is formed by the three components of the crystal momentum, and the control space is formed by any further parameters, e.g., mechanical strain of the crystal \cite{Huang2015,Armitage2018,Wang2015,Huber2016,Li2016,Wang2015,Lu2014};
(ii) spin systems in a homogeneous magnetic field, where the three magnetic-field components form the configurational space, and further parameters are the control parameters \cite{Bruno2006,Scherubl2019,Frank2020};
(iii) multi-terminal Josephson junctions, controlled by more than three parameters such as magnetic fluxes and electric voltages, etc. \cite{Riwar2016,Fatemi2021,Frank2021}.

The central object of our study is the \emph{Weyl phase diagram}: the phase diagram in the control space that shows the number of Weyl points in the configurational space. 
By connecting results of singularity theory with parameter-dependent Hermitian matrices, we find that under certain conditions, Weyl phase diagrams exhibit universal geometrical features, which correspond to the known singularities of generic mappings between manifolds of equal dimension. 

We exemplify this general observation using two example physical setups.
First, we show that the swallowtail singularity, characteristic of mappings between manifolds of dimension $m=3$, appear in the Weyl phase diagram of multi-terminal Weyl Josephson junctions. 
Second, we illustrate the universality of our observation by zero-energy Weyl phase diagrams of class D matrices with random parametrization.
This latter model describes the excitation spectrum of hybrid normal-superconducting systems in the presence of spin-orbit interaction and in the absence of time-reversal symmetry, and the corresponding zero-energy Weyl points appear in a 1D configurational space. 
The numerically obtained zero-energy Weyl phase diagrams exhibit fold lines and cusp points, as expected from our observation that this setting is related to singularities of mappings between 2D manifolds.

The rest of this paper is structured as follows. 
In section II., we summarize those key concepts and results from singularity which we will use to analyse the geometrical features of Weyl phase diagrams.
In section III., we showcase the appearance of the swallowtail singularity, characteristic of maps between 3D manifolds, in the Weyl phase diagram of a Weyl Josephson junction.
In section IV., we visually illustrate the appearance of fold and cusp singularities in the zero-energy Weyl phase diagram of parameter-dependent class D Hamiltonians that describe hybrid normal-superconducting systems. 
Finally, in sections V. and VI., we extend the discussion of our results, and provide conclusions.

\begin{table*}
	\centering
	\begin{tabular}{|c|c|c|c|}
		\hline dim Ct & min dim Cf & Name & Canonical form\\
        \hline
        \hline
		1&1&fold point & $(x^2)$\\
        \hline
		2&1&fold line & $(x^2,y)$\\
        2&1&cusp point & $(x^3+xy,y)$\\
        \hline
        3&1&fold surface & $(x^2,y,z)$\\
        3&1&cusp line & $(x^3+xy,y,z)$\\
        3&1&swallowtail point & $(x^4+x^2y+xz,y,z)$\\
		\hline
        4&1&fold hypersurface  & $(x^2,y,z,w)$\\
        4&1&cusp surface & $(x^3+xy,y,z,w)$\\
        4&1&swallowtail line & $(x^4+x^2y+xz,y,z,w)$\\
        4&1&butterfly point & $(x^5+x^3y+x^2z+xw,y,z,w)$\\
        4&2&elliptic umbilic point  & $(x^2-y^2+xz+yw,xy,z,w)$\\
        4&2&hyperbolic umbilic point & $(x^2+y^2+xz+yw,xy,z,w)$\\
		\hline
	\end{tabular}
	\caption{Singularities of mappings between manifolds of equal dimension $m\leq 4$.
    `Name' and `Canonical form' originate from singularity theory.
    A given singularity can appear on a system's Weyl phase diagram if the system has 
    $\textrm{dim} \, \textrm{Ct}$ control parameters and at least 
    $\text{min} \, \textrm{dim} \, \textrm{Cf}$ configurational parameters. 
    For example, the fold point singularity can appear in the Weyl phase diagram if the system is described by a parameter-dependent Hermitian matrix (implying a 3D configurational space) with a single control parameter.
    In contrast, the elliptic and hyperbolic umbilic points cannot appear on the zero-energy Weyl phase diagram of class-D matrices (1D configurational space) with 4 control parameters, since the configurational space dimension is less than $\textrm{min} \, \textrm{dim} \, \textrm{Cf} = 2$.
 \label{tab:singularity}}
\end{table*}

\section{Singularity theory predicts generic and stable features of Weyl phase diagrams}
\label{sec:singularity}

In this section, we first introduce the key concepts and relations from singularity theory that are relevant for the analysis of Weyl phase diagrams. 
We do this via simple and instructive examples of mappings between manifolds of equal dimension, for dimensions $m=1$, $m=2$, and $m=3$.
Then, we outline the connection between these mathematical concepts and results, and Weyl points and Weyl phase diagrams. 

\subsection{Math example for $m=1$: fold}

\emph{The source manifold.}
Consider the 1D manifold $M^1: \{(3x-x^3, x)|x\in \mathbb{R}\} \in \mathbb{R}^2$, i.e., the graph of a cubic polynomial.

\emph{The projection map.}
We define the projection map $\pi$ such that it maps each point $(t,x)$ of $M^1$ to the first coordinate $t$ of the point. 
That is, $\pi$ is a $M^1 \to \mathbb{R}$ map, i.e., a map between two 1D manifolds.

\emph{The counting function of pre-images.}
To each point $t$ of the codomain of the projection map $\pi$, we can associate the number of pre-images $\#\pi^{-1}(t)$ of that point; we will use $N: \mathbb{R} \to \mathbb{Z}_0^+$ to denote this function, and call it the `counting function of pre-images'. 

\emph{Pre-image phase diagram.}
The function $N$ partitions the codomain of $\pi$. 
There are three partitions that are regions with non-zero length;  these are $]-\infty,-2[$, 
$]-2,2[$, and
$]2,\infty[$, and $N$ takes the value 1, 3, and 1, respectively, in these regions. 
Furthermore, there are two isolated points, $-2$ and $2$, that separate the above regions.
The counting function $N$ takes the value 2 in these points. 
 
The isolated points separating the extended regions are locations of pairwise `creation' or `annihilation' processes of pre-images. 
Let us follow the points of a curve in the target manifold $\mathbb{R}$ from $t<2$ to $t>2$:
as $t$ increases in the range $t<2$, there are 3 pre-images in the source manifold that move, two of them merge to a single point when $t=2$, and those two pre-images disappear (`pairwise annihilation') for $t>2$ where the pre-image count is 1. 

Following physics terminology, we call the extended regions `pre-image phases' or `phases' for short, 
and the isolated points separating them we term `pre-image phase boundaries', or `phase boundaries' for short. 

\emph{Phase boundaries are formed by the singular values of the projection map.}
For the projection map $\pi$, the points of the domain can be classified as regular or singular. 
Regular (singular) points are those where the derivative of the map is non-zero (zero). 
This classification of the points of the domain of $\pi$ is strongly related to the pre-image phase diagram. 
In fact, the images of the singular points of the domain (i.e., the singular values of the map) appear in the pre-image phase diagram as phase boundaries. 

\emph{Extension from the example to generic maps.}
The above picture, although described for the case of a single example, extends naturally to generic maps between 1D manifolds. 
Furthermore, for generic maps between 1D manifolds, the local behavior of the map in any two regular (singular) points is equivalent, in the following sense:
In a regular (singular) point, in appropriately chosen coordinates, the map can be written as $f(x) = x$ ($f(x) = x^2$).
These singular points are also called `fold points' (see Table \ref{tab:singularity}).
Furthermore, for generic maps, the structure of singular points is robust against small deformations of the map, which implies that the pre-image phase diagram is also robust against such small deformations.

\subsection{Math example for $m=2$: cusp}

\emph{The source manifold.}
Consider now the 2D manifold $M^2: \{(t_1, -x^3-t_1x,x)|(x,t_1)\in \mathbb{R}^2\} \in \mathbb{R}^3$, as shown in Fig.~\ref{fig:projection}.

\emph{The projection map.}
We define the projection map $\pi$ such that it maps each point $(t_1, t_2, x)$ of $M^2$ to the first two coordinates $(t_1,t_2)$. 
That is, $\pi$ is a $M^2 \to \mathbb{R}^2$ map, i.e., a map between two 2D manifolds. 
The projection map $\pi$ is also illustrated in Fig.~\ref{fig:projection}.

\emph{Counting function of pre-images.}
To each point $(t_1,t_2)$ of the codomain of the projection map $\pi$, we can associate the number of pre-images $\#\pi^{-1}(t_1,t_2)$ of that point; we will use $N: \mathbb{R}^2 \to \mathbb{Z}_0^+$ to denote this function. 
We call $N$ the `counting function of pre-images'. 

\emph{Pre-image phase diagram.}
The function $N$ partitions the codomain of $\pi$, as illustrated in Fig.~\ref{fig:projection} as the patterns on the $(t_1,t_2)$ plane.
The light gray and dark gray partitions are extended regions with non-zero area, corresponding to pre-image counts of 1 and 3, respectively.
The red curves separating the grey regions correspond to pre-image count of 2, except the cusp point where the two curves meet, corresponding to pre-image count of 1. 

The curve-type boundaries correspond to pairwise `creation' or `annihilation' processes of pre-images.
In fact, the left (right) curve boundary corresponds to the creation or annihilation of the upper (lower) two pre-images. 
The cusp point is a location of a three-point process \cite{Konye2021,Guba2023}, where the number of pre-images change from 1 to 3 such that the two newborn pre-images are created at the position of the original single pre-image. 
Analogously to the $m=1$ case, we call the extended regions with non-zero area `pre-image phases' or 'phases' for short, and the boundaries separating these regions we call `phase boundaries'. 

\emph{Phase boundaries are formed by the singular values of the projection map.}
For the projection map $\pi$, the points of its domain can be classified as regular or singular. 
Regular (singular) points are those where the Jacobian of the map has a non-vanishing (vanishing) determinant. 
Singular points can be further classified, as fold points or cusp points. 
This classification of the points of the domain of $\pi$ is strongly related to the pre-image phase diagram shown in the $(t_1,t_2)$ plane of  Fig.~\ref{fig:projection}:
The images of the fold points of $\pi$ form the curved phase boundary lines, whereas the image of the single cusp point of $\pi$ is the meeting point of the curved phase boundary lines. 

\emph{Extension from the example to generic maps.}
The above picture extends naturally to generic maps between 2D manifolds.
According to Whitney's theorem, singular points of such generic mappings are either cusp points, or fold points forming lines (`fold lines') (see Table \ref{tab:singularity}).
Furthermore, for generic maps between 2D manifolds, the local behavior of the map in all regular [fold] [[cusp]] points is equivalent, in the sense that in appropriately chosen coordinates, the map can be written in the canonical form 
$f(x,y) = (x,y)$ 
[$f(x,y) = (x^2,y)$]
[[$f(x,y) = (x^3 + xy, y)$]].
Furthermore, for generic maps, the structure of singular points is robust against small deformations of the map, which implies that the pre-image phase diagram is also robust against such deformations. 

\subsection{Math example, $m=3$:  swallowtail}

\emph{The source manifold.}
Consider now the 3D manifold $M^3: \{(t_1,t_2, -x^4-t_1x^2-t_2x,x)|(t_1,t_2,x)\in \mathbb{R}^3\} \in \mathbb{R}^4$.

\emph{The projection map.}
We define the projection map $\pi$ such that it maps each point $(t_1,t_2,t_3,x)$ of $M^3$ to the first three coordinates $(t_1,t_2,t_3)$.
That is, $\pi$ is a \mbox{$M^3 \to \mathbb{R}^3$} map, i.e., a map between two 3D manifolds. 

\emph{Counting function of pre-images.}
To each point $(t_1,t_2,t_3)$ of the codomain of the projection map $\pi$, we can associate the number of pre-images $\#\pi^{-1}(t_1,t_2,t_3)$ of that point; we will use $N: \mathbb{R}^3 \to \mathbb{Z}_0^+$ to denote this function, and call it the `counting function of pre-images'. 

\emph{Pre-image phase diagram.}
The function $N$ partitions the codomain of $\pi$. 
This partitioning is shown in Fig.~\ref{fig:WJC}b.
As illustrated there, there are extended partitions of non-zero volume, there are surfaces (fold surfaces) that separate the regions, there are two curves (cusp lines) that separate the surfaces, there is an intersection curve of the fold surfaces, and there is a single point (swallowtail point), where the curves meet. 
The counting function takes the values 0, 2, and 4, in the bottom, top, and middle regions of the figure.
Along the fold surfaces, the pre-image count is 3. 
Along the intersection curve of the two fold surfaces, it is 2. 
Along the cusp lines, it is also 2. 
In the swallowtail point, it is 1.

The fold surface phase boundaries correspond to pairwise creation or annihilation of pre-images. 
The cusp lines correspond to `three-point processes' \cite{Konye2021,Guba2023}, where three pre-images merge into a single one.
The intersection curve of the fold surfaces corresponds to `simultaneous two-point processes', where the four pre-images merge and annihilate in two pairs, simultaneously.
The swallowtail point corresponds to a `four-point process', where the four pre-images merge in a single location and annihilate. 
We call the extended regions with non-zero volume `phases', and the boundaries separating these regions `phase boundaries'. 

\emph{Phase boundaries are formed by the singular values of the projection map.}
For the projection map $\pi$, the points of its domain can be classified as regular or singular. 
Regular (singular) points are those where the Jacobian of the map has a non-vanishing (vanishing) determinant. 
Singular points can be further classified, as fold points, cusp points, or swallowtail points.
This classification of the points of the domain of $\pi$ is strongly related to the pre-image phase diagram shown in Fig.~\ref{fig:WJC}b:
The image of the surface formed by the fold points of $\pi$ form the fold surfaces in Fig.~\ref{fig:WJC}b, the images of the curves of the cusp points of $\pi$ form the cusp lines in Fig.~\ref{fig:WJC}b, and the image of the single swallowtail point of $\pi$ is the swallowtail point in Fig.~\ref{fig:WJC}b. 

\emph{Extension from the example to generic maps.}
The above picture extends naturally to generic maps between 3D manifolds. 
Singular points of such maps are either swallowtail points, or cusp points forming lines (`cusp lines'), or fold points forming surfaces (`fold surfaces'), see Table \ref{tab:singularity}.
Furthermore, for generic maps between 3D manifolds, the local behavior of the map in any two regular [fold] [[cusp]] [[[swallowtail]]] point is equivalent, in the sense that in appropriately chosen coordinates, the map can be written in the canonical form $f(x,y,z) = (x,y,z)$ [$f(x,y,z) = (x^2,y,z)$] [[$f(x,y,z) = (x^3+xy,y,z)$]] [[[$f(x,y,z) = (x^4+x^2y+xz,y,z)$]]].
Furthermore, for generic maps, the structure of singular points is robust against small deformations of the map, which implies that the pre-image phase diagram is also robust against such deformations.

\subsection{Weyl phase diagrams}
\label{subsec:weylphasediagrams}

In this work, we focus on physical systems that are described by parameter-dependent Hamiltonians, i.e., Hermitian matrices.
In particular, we assume that the number of parameters $n_\text p$ is at least 4, and the parameters are naturally grouped into two groups, of size 3 (configurational parameters) and $m = n_\text p -3$ (control parameters).
We denote the configurational space as $\textrm{Cf}^3$ and the control space as $\textrm{Ct}^m$.

For a fixed set of the control parameters, the energy eigenvalues as functions of configurational parameters (`energy bands') might exhibit generic twofold degeneracies (Weyl points) or more exotic degeneracy patterns \cite{Neumann1929,Herring1937}.
Focus our attention to degeneracies between two specific bands -- without the loss of generality, let us choose the bands of the ground state and the first excited state. 
As the control parameters are varied continuously, the degeneracy points `evolve’: generically, the Weyl points of the two lowest-energy bands move in the configurational space, and for special values of the control parameters, Weyl points can merge and annihilate, or Weyl points can be created. Control parameter values where Weyl points are created or annihiliated are regarded as `phase boundaries’, separating different regions (`phases’) in the control space  characterized by different numbers of Weyl points.
We call this partitioning of the control parameter space a `Weyl phase diagram'. 

Next, we argue that the Weyl phase diagram is actually a special case of a pre-image phase diagram, introduced in the previous subsections for $m=1,2,3$.
What is the corresponding source manifold, projection map, and target manifold?
The source manifold is the `surface' $\textrm{W}^m \subset \textrm{Cf}^3 \times \textrm{Ct}^m$ drawn by the Weyl points in the product of the configuration space and the control space. 
Recall that Weyl points are isolated points (i.e., zero-dimensional objects) in the configurational space, and the product of the configuration space and the control space is $(3+m)$-dimensional, hence the Weyl points draw an $m$-dimensional manifold $\textrm{W}^m$ in the product space. 

The projection map $\pi: \textrm{W}^m \to \textrm{Ct}^m, (k,t) \mapsto t$ is defined as the projection from the $m$-manifold of Weyl points on the control space.
The counting function of pre-images $N$, defined in the previous subsections, can be also defined for this projection map $\pi$, and the corresponding pre-image phase diagram provides the Weyl phase diagram. 

To conclude, we found that under generic conditions, a Weyl phase diagram of dimension $m$ is a pre-image phase diagram of a specific projection map, and hence its geometric features are universal: the phase diagram consists of extended regions (phases) where the number of Weyl points is constant, these phases are separated by phase boundaries formed by the singular values of the projection map, and these phase-boundary points carry universal geometrical characteristics determined by their singularity class. 
In particular, for $m=2$, the phase boundary consists of fold lines that may meet at cusp points, and for $m=3$, the phase boundary consists of fold surfaces, that may meet in cusp lines, that may meet in swallowtail points. 
We note that the list of singularities is enriched further as $m$ increases above 3, as exemplified in the lowest block of Table \ref{tab:singularity}.

\section{Swallowtail singularity in a Weyl--Josephson circuit}
\label{sec:WJC}

To demonstrate the Weyl-point singularities in a concrete physical system, we consider the Weyl Josephson circuit, originally proposed in Fig.~1 of \cite{Fatemi2021}. The circuit consists of superconductor islands connected by Josephson junctions which form loops (Fig.~\ref{fig:WJC}a).
In this setup, Weyl points are defined in a 3D configurational space (fluxes), the 3D control space consists of gate-voltage parameters, and the singularities (fold surfaces, cusp lines, and the swallowtail point) appear in the 3D Weyl phase diagram defined in the control (gate-voltage) parameter space. 

\subsection{Hamiltonian}
The Hamiltonian of the circuit reads
\begin{eqnarray}\label{eq:WJC}
\hat{H}(\boldsymbol \varphi,\vec n_\text g) &=&
E_{\text C}
\left(\hat{\vec n}-\vec n_{\text g}\right)
\cdot 
c^{-1}
\left(\hat{\vec n}-\vec n_{\text g}\right)\\
&-& 
\sum\limits_{\substack{\alpha,\beta=0 \\ \alpha<\beta}}^{3} 
E_{\text{J}, \alpha \beta} 
\cos\left[\hat{\varphi}_\alpha-\hat{\varphi}_\beta+\gamma_{\alpha \beta}(\varphi_x, \varphi_y,\varphi_z)\right].\nonumber
\end{eqnarray}
The first term in the Hamiltonian is the charging energy term where the charging energy scale $E_{\text C}=(2e)^2/(2C_0)\approx 77.5\text{ GHz}$ is set by the capacitance scale $C_0=1\text{ fF}$, and $c=C/C_0$ is the dimensionless capacitance matrix defined from the capacitance matrix \cite{vanderWielRMP} $C$ of the circuit. The elements of the vector $\hat{\vec n} = (\hat{n}_1,\hat{n}_2,\hat{n}_3)$ are the number operators $\hat{n}_\alpha$ counting the Cooper pairs on the islands $\alpha \in \{1,2,3\}$. The gate voltage $V_{\text{g},\alpha}$ coupled to the $\alpha$th island through the capacitance $C_{\text{g},\alpha}$ shifts the number operator in the Hamiltonian by the effective offset charge $n_{\text{g},\alpha}=C_{\text{g},\alpha}V_{\text{g},\alpha}/(2e)$. 

The second term in the Hamiltonian is the tunneling term with the Josephson energies $E_{\text{J}, \alpha \beta}$ of the junctions between island $\alpha$ and $\beta$, with the phase operators $\hat\varphi_i$ canonically conjugated to the number operators $\hat{n}_i$. The control angles $\gamma_{\alpha \beta}$ are given by
$\gamma_{0\beta}=0$,
$\gamma_{12}=\varphi_x$,
\mbox{$\gamma_{13}=-\varphi_z$},
and $\gamma_{23}=\varphi_y$
with the magnetic fluxes $\varphi_i=\pi\Phi_i/\Phi_0$ of the loops. The Josephson energies and capacitances are given in Table~\ref{tab:WJC}.

The Hamiltonian is truncated to the 8-dimensional subspace spanned by the number operator eigenstates $\ket{n_1,n_2,n_3}$ with $n_i\in\{0,1\}$. Degeneracies between the ground state and first excited state is investigated. The magnetic fluxes and offset charges gives $n_\text p=6$-dimensional total parameter space divided into $3+3$, where we choose the magnetic fluxes to be the configurational parameters hosting the Weyl points.

\begin{table}[h!]
	\centering
	\begin{tabular}{|c|cccccc|}
		\hline $\alpha \beta$ &01&02&03&12&13&23\\
        \hline
		$E_{\text{J}, \alpha \beta}$ (GHz) &2&4&6&3&3&6\\
		$C_{\alpha \beta}$ (fF) &2&1&2&3&4&3\\
		\hline
	\end{tabular}
	\caption{
	Weyl--Josephson circuit parameters used in the numerical calculations yielding Fig.~\ref{fig:WJC}c-k. 
    Gate capacitances are set to \mbox{$C_{\text{g},1} = C_{\text{g},2} = C_{\text{g},3} = 0.1\, \text{fF}$. \label{tab:WJC}}}
\end{table}

\subsection{Symmetries}
The Hamiltonian has an effective time-reversal and inversion symmetry
\begin{eqnarray}
    H(-\boldsymbol \varphi,\vec n_\text g)&=&H^\ast(\boldsymbol \varphi,\vec n_\text g)\label{eq:WJCsymm}, \\
    H(-\boldsymbol \varphi,1-\vec n_\text g)&=&PH(\boldsymbol \varphi,\vec n_\text g)P^{-1}, \label{eq:WJCsymm2}
\end{eqnarray}
with $P\ket{n_1,n_2,n_3}=\ket{1-n_1,1-n_2,1-n_3}$. The consequence of  Eq.~\eqref{eq:WJCsymm} is that $H(\boldsymbol \varphi,\vec n_\text g)$ and $H(-\boldsymbol \varphi,\vec n_\text g)$ has the same spectra, meaning that a Weyl point located at $\boldsymbol{\varphi}_\text{WP}$ has a time-reversal partner with the same chirality at $-\boldsymbol{\varphi}_\text{WP}$ for any $\vec n_{\text g}$. The two symmetries together results that $H(\boldsymbol{\varphi},\bold{1/2})=PH^{\ast}(\boldsymbol{\varphi},\bold{1/2})P^{-1}$ with $\bold{1/2}:=(1/2,1/2,1/2)$ for any $\boldsymbol{\varphi}$, meaning that it is possible to do a constant (not depending on $\boldsymbol{\varphi}$) basis transformation so that $U H(\boldsymbol{\varphi},\bold{1/2})U^{-1}$ is a real-valued matrix. This lowers the codimension of the band crossings to 2 in the special control point $\vec n_\text g = \bold{1/2}$, meaning that the general degeneracy pattern in the 3-dimensional configurational space is a 1-dimensional nodal loop \cite{Fatemi2021,Frank2021}.

Due to the periodicity of the configurational (flux) parameter space, the total topological charge, i.e., the sum of topological charges of all the Weyl points is zero \cite{Nielsen1981}. Therefore, the number of  Weyl points must be even. 
Due to the additional conditions that (1) Weyl points come in time-reversal pairs, and (2) the two Weyl points of a time-reversed pair carry the same topological charge, the number of Weyl points must be a multiple of 4. 

\subsection{Weyl points}
To investigate exotic Weyl-point merging processes one needs as many Weyl points in the configurational space as possible. To achieve this we search the Weyl points in the vicinity of the nodal loop control parameter point \mbox{$\vec n_\text {g,loop}=\bold{1/2}$}. This is advantageous as the nodal loop can be used as source of Weyl points. Upon a small perturbation $\vec n_\text {g,loop}+\delta \vec n_\text {g}$ the nodal loop breaks into multiple Weyl points. The perturbation $\vec n_\text {g,loop}+t\vec e$ in the direction $\vec e=(-4,1,9)/\sqrt{98}$ results 8 Weyl points (4 time-reversal symmetric Weyl point pairs) with sufficiently small $t$. For larger $t$ the 8-point region curves away from the straight line.

\begin{figure*}
	\begin{center}
		\includegraphics[width=2\columnwidth]{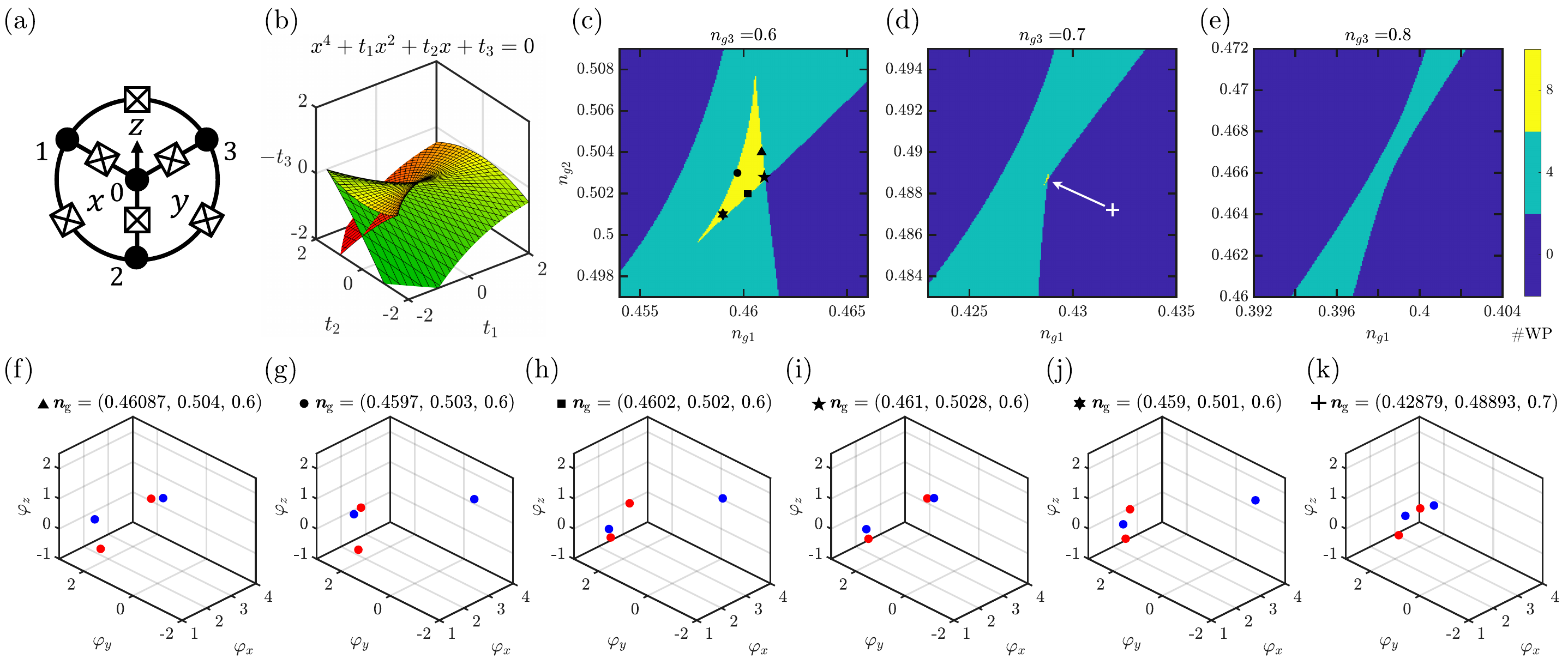}
  \end{center}
\caption{Swallowtail singularity in the Weyl--Josephson circuit. (a) The layout of the Weyl--Josephson circuit  The system can be tuned with changing the magnetic fluxes $\boldsymbol{\varphi}$ through the loops or the voltage differences (offset charges $\vec n_\text g$) between the superconducting islands. Every other parameters such that the Josephson energies and capacitances are set to be constant. (b) Swallowtail singularity illustrated with the roots of the depressed quartic equation $x^4+t_1x^2+t_2x+t_3=0$. The merger of (real valued) roots correspond to the characteristic self-intersecting surface in the 3D control parameter space of coefficients. (c-e) 2D cuts of the Weyl phase diagram of control parameters showing the number of Weyl points in the configurational parameter space revealing a similar structure. The triangular 8-point region disappears as $n_{\text{g}3}$ increased. The generic boundaries between the regions are fold lines which can cross each other and also touch at cusp points. These cusp points form two cusp lines in the 3D Weyl phase diagram which touch at the swallowtail point. The possible Weyl-point mergers in the configurational space correspond to the points marked in the Weyl phase diagram are shown in panel (f-k). The points are colored by their topological charge: red (blue) points have charge +1 (-1).
}
\label{fig:WJC}
\end{figure*}

Fig.~\ref{fig:WJC}c-e show 2D cuts of the Weyl phase diagram which reveal the characteristic shape of a swallowtail singularity corresponding to the interaction of 4 Weyl points with alternating topological charges (the time-reversal pairs are far). In Fig.~\ref{fig:WJC}c for $n_{\text g,3}=0.6$ the 8-point region (yellow) appears with a triangular shape with 3 different boundaries, in Fig.~\ref{fig:WJC}d this triangle shrinks, and in Fig.~\ref{fig:WJC}e it is absent. This corresponds to the 2D $(t_2,t_3)$ cuts of the swallowtail shown in Fig.~\ref{fig:WJC}b. The boundaries are fold lines, which correspond to the merger and annihilation of 2 oppositely charged Weyl points (see Fig.~\ref{fig:WJC}f-h). This can happen between the 2 leftmost, between the 2 middle, or between the 2 rightmost points. The merger of the 2 leftmost and the 2 rightmost Weyl points are independent, hence these fold lines intersect at a point, where the two mergers coincide (Fig.~\ref{fig:WJC}i). The merger of the 2 middle points and the merger of the 2 leftmost (rightmost) points are not independent, their fold lines are touching each other at a cusp point, which correspond to the merger of the 3 leftmost (rightmost) points, see Fig.~\ref{fig:WJC}j. The Weyl phase diagram is actually three-dimensional with fold surfaces and cusp lines. The two cusp lines touch each other at the swallowtail point where the 4 Weyl points merge at a single point. This is illustrated in Fig.~\ref{fig:WJC}d where the triangular 8-point region is almost disappeared and in Fig.~\ref{fig:WJC}k, where the corresponding Weyl point configuration at the '+' marker shows 4 Weyl points close together. We found that the actual swallowtail point is at $\vec n_{\text{g,swallowtail}}=(0.418,0.481,0.735)$.

\section{Cusp and fold singularities in superconducting systems of class D}
\label{sec:classD}

In the preceding sections, we focused on parameter-dependent $n \times n$ Hermitian matrices with a 3D configurational space and an $m$-dimensional control space. 
Our  considerations above, concerning the Weyl points of these matrices, hold for any pair of neighboring bands $(j,j+1)$, where $1 \leq j \leq n-1$.
For these Hermitian matrices, the Weyl points are zero-dimensional objects in the 3D configuration space. 

There are many quantum-mechanical models where the Hamiltonian is not a generic Hermitian matrix, but a constrained one.
In particular, the tenfold-way classification of Altland and Zirnbauer defines Hamiltonian classes constrained by various combinations of time-reversal, particle-hole, and chiral symmetries \cite{Altland1997}.
In this section, we present our results corresponding to the Altland-Zirnbauer class D, which represents Hamiltonians (also called Bogoliubov--de Gennes Hamiltonians or BdG Hamiltonians) describing excitations in superconductors or hybrid normal-superconductor hybrid systems.
A typical setup modelled by matrices of class D is a (possibly multi-terminal) Josephson junction in the presence of spin-orbit interaction and time-reversal-breaking magnetic fields, and in the absence of charging effects \cite{Chtchelkatchev2003,Hanho2022thesis,Coraiola2023}.
Non-interacting models of one-dimensional topological superconductors hosting Majorana zero modes also fall into class D \cite{Beenakker2015}.

Studying the properties of parameter-dependent class-D matrices is  motivated by the intense experimental efforts on superconducting devices modelled by such matrices. 
However, here we focus on class D also because certain aspects of the  singularity-theory analysis of their Weyl points can be visualised in a particularly straightforward manner using surface plots.
To appreciate this, we first note that class D matrices have even dimension, i.e., $n = 2 n_s$ with $n_s$ being a positive integer.
Furthermore, the eigenvalue spectrum of class D matrices is symmetric with respect to zero. 
Finally, the generic eigenvalue degeneracies between bands $n_s$ and $n_s+1$, which necessarily happen at zero energy, and sometimes referred to as `parity switches', are special in the sense that they appear for \emph{single-parameter} families of matrices, as opposed to Weyl points of Hermitian matrices which require three parameters to be varied.
In what follows, we will use the term `zero-energy Weyl points' for parity switches of single-parameter class D matrix families. 

Consider now a physical system described by a parameter-dependent class-D matrix, where the number of parameters is $n_\text p = 3$, which are grouped into a single-parameter group defining the 1D configurational space and the two other parameters forming the control space of dimension $m=2$.  
We might be interested in the number of zero-energy Weyl points in the configurational space, and how that number changes as the control parameters are varied. 
This dependence is characterized by the zero-energy Weyl phase diagram.
This zero-energy Weyl phase diagram has certain universal geometric properties, which follows from Whitney's theorem describing the singularities of mappings between 2D manifolds.
Namely, the zero-energy Weyl phase diagram consists of extended regions of finite area where the number of zero-energy Weyl points is constant, and phase boundaries constructed from fold lines that might meet in cusp points.

We now illustrate these universal geometric properties using a random-matrix approach. The BdG Hamiltonian depends on 3 parameters in the following way:
\begin{eqnarray}\label{eq:H_BdG}
H(\alpha,\beta,\gamma)&=&H_1\cos(\alpha)\cos(\beta)\cos(\gamma)\nonumber\\
&+&H_2\cos(\alpha)\cos(\beta)\sin(\gamma)\nonumber\\
&+&H_3\cos(\alpha)\sin(\beta)\cos(\gamma)\nonumber\\
&+&H_4\cos(\alpha)\sin(\beta)\sin(\gamma)\nonumber\\
&+&H_5\sin(\alpha)\cos(\beta)\cos(\gamma)\\
&+&H_6\sin(\alpha)\cos(\beta)\sin(\gamma)\nonumber\\
&+&H_7\sin(\alpha)\sin(\beta)\cos(\gamma)\nonumber\\
&+&H_8\sin(\alpha)\sin(\beta)\sin(\gamma),\nonumber
\end{eqnarray}
where $H_n$ are random matrices with the structure
\begin{eqnarray}\label{eq:H_BdG_block}
    H_n=\begin{pmatrix}
        H_{0,n}&\Delta_n\\
        -\Delta_n^\ast&-H_{0,n}^\ast
    \end{pmatrix},
\end{eqnarray}
where $\Delta_n$ is a skew-symmetric complex matrix and $H_{0,n}$ is Hermitian. We constructed these matrices with \mbox{$\Delta_n=d_n-d_n^{\text T}$} and \mbox{$H_{0,n}=h_n+h_n^{\dagger}$} where the entries are pseudo-random numbers between -1/2 and 1/2, defined via
\begin{eqnarray}\label{eq:H_BdG_entries}
\Re d_{n,kl}&=&\left\{\sqrt{2}k+\sqrt{3}l+\sqrt{5}n\right\}-\frac{1}{2}\\
\Im d_{n,kl}&=&\left\{\sqrt{6}k+\sqrt{7}l+\sqrt{10}n\right\}-\frac{1}{2}\\
\Re h_{n,kl}&=&\left\{\sqrt{11}k+\sqrt{13}l+\sqrt{14}n\right\}-\frac{1}{2}\\
\Im h_{n,kl}&=&\left\{\sqrt{15}k+\sqrt{17}l+\sqrt{19}n\right\}-\frac{1}{2},
\end{eqnarray}
where $\{x\}$ denotes the fractional part of $x$. In this example the dimension of the full BdG Hamiltonian is $n=12\times 12$.

The BdG Hamiltonian is skew-symmetric in the Majorana basis, resulting a symmetric spectrum. It also has a so-called Pfaffian for which $\text{pf}(H)^2=\det(H)$. The Pfaffian is a polynomial of the entries of the matrix. It changes sign when two energy levels cross at zero. Therefore, zero-energy degeneracies appear with the fine-tuning of only 1 parameter. In a 3D parameter space they generally form a 2D manifold. 

\begin{figure}
	\begin{center}
		\includegraphics[width=0.9\columnwidth]{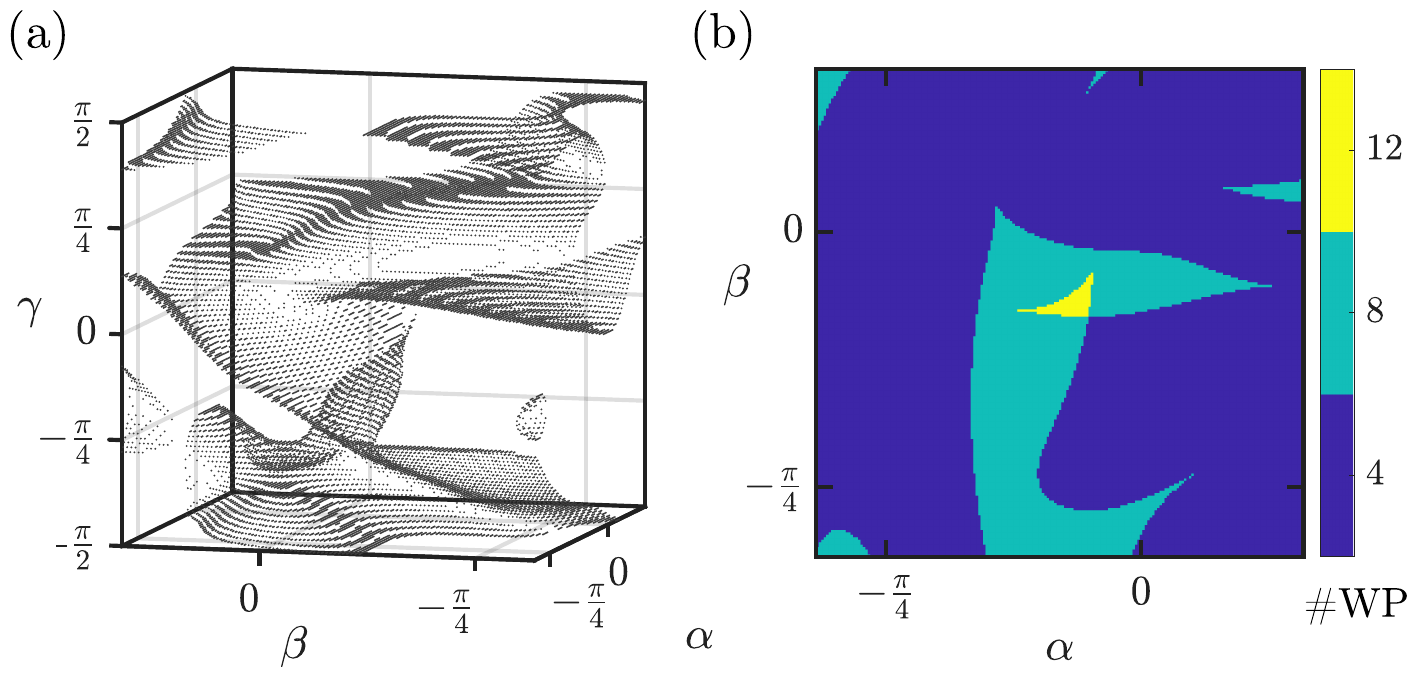}
  \end{center}
\caption{Fold lines and cusp points as singularities in the zero-energy Weyl phase diagram of a random BdG Hamiltonian. (a) Due to the particle-hole symmetry, zero-energy degeneracies appear with the fine-tuning of a single parameter. Therefore, zero-energy degeneracies appear as surfaces in a 3D parameter space. (b) Zero-energy Weyl phase diagram corresponding to the vertical projection of the surface on (a), i.e., obtained by counting the degeneracy points along the vertical direction. The phase diagram exhibits the generic and robust singularities in 2D: fold lines and cusp points.}
\label{fig:BdG}
\end{figure}

Fig.~\ref{fig:BdG}a shows the zero-energy  degeneracy surface of the pseudo-random BdG Hamiltonian in the total parameter space. The figure is produced with calculating the Pfaffian on a $100\times100\times100$ grid and highlighting points where the Pfaffian changes sign. We divided the total parameter space into the configurational parameter $\gamma$ and to the control parameters $(\alpha,\beta)$. For a fixed $\alpha$ and $\beta$, we counted the sign changes of the Pfaffian along the $\gamma$ axis; plotting these counts as the function of $\alpha$ and $\beta$ provides the zero-energy Weyl phase diagram as shown in Fig.~\ref{fig:BdG}b. We created this phase diagram using $200\times200\times200$ grid.

The parameter dependence of the Hamiltonian in Eq.~\eqref{eq:H_BdG} is given in a way that by shifting any angle by $\pi$ results in the negative of the Hamiltonian, e.g., \mbox{$H(\alpha,\beta,\gamma+\pi)=-H(\alpha,\beta,\gamma)$}. Because the negative have the same spectrum, the Weyl phase diagram is $\pi$ periodic and in the generic points the Weyl-point number is divisible by 4.

The zero-energy Weyl phase diagram of the BdG Hamiltonian shows a rich structure with the stable singularities of 2D manifolds: fold lines meeting in cusp points and crossing each other. Fig.~\ref{fig:BdG} highlights the interval $-1\leq\alpha,\beta\leq1/2$ which resembles a 2D cut of the phase diagram of a swallowtail singularity. An additional angle parameter might complete the swallowtail singularity if the two cusp lines meet upon changing the new parameter. The total phase diagram is crowded with the singularities. This structure of singularities becomes more complicated upon increasing the dimension of the Hilbert space (not shown) because this also increases the number of zero-energy Weyl points in the configurational space, leading to more possible mergers between them.

\section{Discussion}
\label{sec:discussion}
\subsection{When does the set of Weyl points form a manifold?}

In Sec.~\ref{subsec:weylphasediagrams}, we have argued that the set of Weyl points in the total parameter space $\textrm{Cf}^3 \times \textrm{Ct}^m$ forms a manifold.
Based on this precondition, we highlighted and exploited a strong connection between the Weyl-point merging processes and the stable mappings between manifolds of equal dimension.
We discuss this precondition further in this subsection.

The $n\times n$ Hermitian matrices form a $n^2$-dimensional real vector space. The subset of two-fold degenerate matrices is a $n^2-3$-dimensional (3 codimensional) submanifold \cite{Neumann1929,Herring1937}.
Furthermore, the set of matrices with two-fold degeneracy between the $i$-th and $(i+1)$-th eigenvalues and those with two-fold degeneracy between the $(i+1)$-th and $(i+2)$-th eigenvalues meet at points with a three-fold degeneracy with dimension $n^2-8$. In the following we denote the two-fold degeneracy set between the $i$-th and $(i+1)$-th eigenvalues by $\Sigma$. Note that our arguments remain true for the whole two-fold degeneracy set.

The Hamiltonian of a physical system is map \mbox{$H: \text{Cf}^3\times\text{Ct}^m\to \text{Herm}(n)$} from the total parameter space to the space of $n\times n$ Hermitian matrices. 
The set of Weyl points corresponding to the two-fold degeneracy set between the $i$-th and $(i+1)$-th eigenvalues is the pre-image $H^{-1}(\Sigma)$. According to the transversality theorem \cite{Thom1952}, a generic Hamiltonian map $H$ is transverse to $\Sigma$ (intuitively, `non-tangential') and the pre-image $H^{-1}(\Sigma)$ is a submanifold in the total parameter space of codimension 3.

Based on the above considerations, we can envision situations when the set of Weyl points is \emph{not} a manifold. 
For example, this is the case when the image of the Hamiltonian map is tangential to the two-fold degeneracy set $\Sigma$, i.e., the intersection is non-generic; or if the image of the Hamiltonian map intersects a multi-fold degeneracy set. 
The former case might arise in case of fine tuning or symmetries, i.e., it does not arise when the mapping is generic. 
The latter case is also non-generic if $n_\text{p} < 8$, e.g., in the case $n_\text{p} = 6$ studied in Sec.~\ref{sec:WJC}.
However, for $n_\text{p}\geq8$, stable intersections of the image of the Hamiltonian map and the multi-fold degeneracy sets can arise, and the whole degeneracy set is not a manifold. In this case, our argument is still valid \emph{locally}, in a small neighbourhood of a two-fold degeneracy with a finite gap from the other levels.

Note also that for our argument a further condition should hold as well, namely, the projection $\pi$ has to be generic. Despite we assumed that $H$ is generic, $\pi$ is not necessarily a generic map. Without providing a full analysis of this condition, we note that if $x\mapsto H(x, t)$ is generic as a deformation of $x\mapsto H(x, t_0)$ for every $t_0$, then the condition is satisfied.

\subsection{Not all singularities appear on Weyl phase boundaries}
In Secs.~\ref{sec:singularity}, \ref{sec:WJC}, and \ref{sec:classD}, we have argued and illustrated that Weyl phase diagrams are pre-image phase diagrams of mappings between manifolds of equal dimension, and that each point of a phase boundary on a Weyl phase diagram belongs to a singularity type. This result raises the following natural question: are all singularity types realised as phase boundary points of Weyl phase diagrams? No — as we show in this subsection.

Sec.~\ref{sec:classD} shows an example where the Hamiltonian has a symmetry which lowers the codimension of a two-fold degeneracy at zero energy to be 1. The corresponding configurational parameter space is therefore 1-dimensional with 1D Weyl points. Similarly, for $\mathcal{PT}$-symmetric Hamiltonians the codimension of a two-fold degeneracy is 2, thus, the configurational space is 2-dimensional with 2D Weyl points. We denote the codimension of the two-fold degeneracy with $0< l\leq3$.

The $n_\text p=l+m$-dimensional total parameter space has an $m$-dimensional Weyl submanifold $\textrm{W}^m$. We defined the projection $\pi: \text{W}^m\to\text{Ct}^m$ as it erases the first $l$ configurational coordinates of the points of the Weyl manifold. Defining the `total projection' $\Pi: \text{Cf}^l\times\text{Ct}^m\to\text{Ct}^m$ with the same definition $\Pi(x,t)= t$ for the total parameter space, we get a mapping with corank $l$, everywhere in the domain of $\Pi$. Restricting the total projection $\Pi$ to $\text{W}^m$ results $\pi=\Pi|_{\text{W}^m}$. Therefore, the mapping $\pi$ has a corank smaller or equal to $l$. Recall that the corank of a map $f$ at a point $w$ of the domain is defined as the corank of the Jacobian matrix $\mbox{Jac}_w(f)$ of $f$ at $w$.  Clearly the corank of $\pi$ at a point $w \in W^m$ is exactly the dimension of the intersection of the $m$-dimensional tangent space $T_w W^m$ of $\text{W}^m$ at $w$ and the $l$-dimensional kernel of the Jacobian $\mbox{Jac}_w(\Pi)$ of $\Pi$ at $w$. Since this corank is at most $l$, in a Weyl-point merging process only those singularities appear whose corank is less than or equal to the dimension $l$ of the dimension of the configurational parameter space.

Concrete examples of `missing singularities’ are the elliptic umbilic point and the hyperbolic umbilic point, listed as the last two entries in Table~\ref{tab:singularity}., which cannot appear as stable features in zero-energy Weyl phase diagrams of class-D systems.

As seen from Table~\ref{tab:singularity}., these singularities are characteristic of generic maps between manifolds of dimension $m=4$, hence it is plausible to search for them in zero-energy Weyl phase diagrams of class-D systems controlled by 5 parameters ($l = 1$ configurational, $m =4$ control). However,  as seen from the corresponding canonical forms in Table~\ref{tab:singularity}., the corank of these singularities is 2. The consideration of the previous paragraph, on the other hand, implies that the corank of the projection map $\pi$ is at most $l$, which is 1 in this case. As a consequence, umbilic points do not appear on zero-energy Weyl phase diagrams of class D systems.

\section{Conclusions}
\label{sec:conclusions}
To conclude, we have argued that singularities of maps between manifolds of equal dimension naturally appear in Weyl phase diagrams of parameter-dependent Hermitian matrices, and illustrated this by numerical results revealing the swallowtail singularity in the gate-voltage control parameter space of a Weyl Josephson junction. We have also illustrated singularities (fold and cusp) on the zero-energy Weyl phase diagram of a parameter-dependent class-D Hermitian matrices, which describe superconducting nanostructures. Based on our arguments, we expect that the results generalise in a broad range of systems; for example, Weyl phase diagrams representing Weyl-point creation and annihilation in electron (phonon, magnon, photon) band structures show similar universal geometrical features characterised by singularities.

\acknowledgments

We thank Z.~Guba for useful discussions. 
This research was supported by the Ministry of Culture and Innovation and the National Research, Development and Innovation Office (NKFIH) within the Quantum Information National Laboratory of Hungary (Grant No. 2022-2.1.1-NL-2022-00004), and by NKFIH via the OTKA Grant No. 132146.

Supported by the ÚNKP-22-3-II-BME-6 New National Excellence Program of the Ministry for Culture and Innovation from the source of the National Research, Development and Innovation Fund.

\bibliography{SingularityTheory.bbl}
\end{document}